\documentclass[reprint,amsmath,amssymb,aps,groupedaddress,superscriptaddress]{revtex4-2}
\usepackage{graphicx}
\usepackage{dcolumn}
\usepackage{bm}
\setcitestyle{super}
\usepackage{caption}
\usepackage{float}
\usepackage{xcolor}
\usepackage{soul}
\usepackage[]{hyperref}
\hypersetup{
  colorlinks   = true, 
  urlcolor     = blue, 
  linkcolor    = blue, 
  citecolor   = blue 
}

\usepackage{MnSymbol}

\begin{document}

\title{Time-resolved Coulomb collision of single electrons}

\author{J.D. Fletcher}
\email{jonathan.fletcher@npl.co.uk}
\affiliation{National Physical Laboratory, Hampton Road, Teddington TW11 0LW, United Kingdom}
\author{W. Park} 
\affiliation{Department of Physics, Korea Advanced Institute of Science and Technology, Daejeon 34141, Korea}
\author{S. Ryu} 
\affiliation{Instituto de F\'{i}sica Interdisciplinary Sistemas Complejos IFISC (CSIC-UIB), E-07122 Palma de Mallorca, Spain}
\author{P. See} 
\affiliation{National Physical Laboratory, Hampton Road, Teddington TW11 0LW, United Kingdom}
\author{J.P. Griffiths} 
\affiliation{Cavendish Laboratory, University of Cambridge, J. J. Thomson Avenue, Cambridge CB3 0HE, United Kingdom}
\author{G.A.C. Jones} 
\affiliation{Cavendish Laboratory, University of Cambridge, J. J. Thomson Avenue, Cambridge CB3 0HE, United Kingdom}
\author{I. Farrer} 
\affiliation{Cavendish Laboratory, University of Cambridge, J. J. Thomson Avenue, Cambridge CB3 0HE, United Kingdom}
\author{D.A. Ritchie} 
\affiliation{Cavendish Laboratory, University of Cambridge, J. J. Thomson Avenue, Cambridge CB3 0HE, United Kingdom}
\author{H.-S. Sim}
\affiliation{Department of Physics, Korea Advanced Institute of Science and Technology, Daejeon 34141, Korea}
\author{M. Kataoka}
\affiliation{National Physical Laboratory, Hampton Road, Teddington TW11 0LW, United Kingdom}

\date{\today} 
\begin{abstract}
\end{abstract}
\maketitle

\textbf{Precise control over interactions between ballistic electrons will enable us to exploit Coulomb interactions in novel ways, to develop high-speed sensing,\cite{Johnson2017} to reach a non-linear regime in electron quantum optics and to realise schemes for fundamental two-qubit operations\cite{Zajac_2018} on flying electrons. Time-resolved collisions between electrons have been used to probe the indistinguishability,\cite{BocquillonScience2013} Wigner function\cite{Jullien-2014-1,Bisognin-2019-2} and decoherence\cite{freulon2015hong} of single electron wavepackets. Due to the effects of screening, none of these experiments were performed in a regime where Coulomb interactions were particularly strong. Here we explore the Coulomb collision of two high energy electrons in counter-propagating ballistic edge states.\cite{fletcher2013clock,ubbelohde2015partitioning} We show that, in this kind of unscreened device, the partitioning probabilities at different electron arrival times and barrier height are shaped by Coulomb repulsion between the electrons. This prevents the wavepacket overlap required for the manifestation of fermionic exchange statistics but suggests a new class of devices for studying and manipulating interactions of ballistic single electrons.
}

\begin{figure}[!ht]
\includegraphics[width=0.3\textwidth]{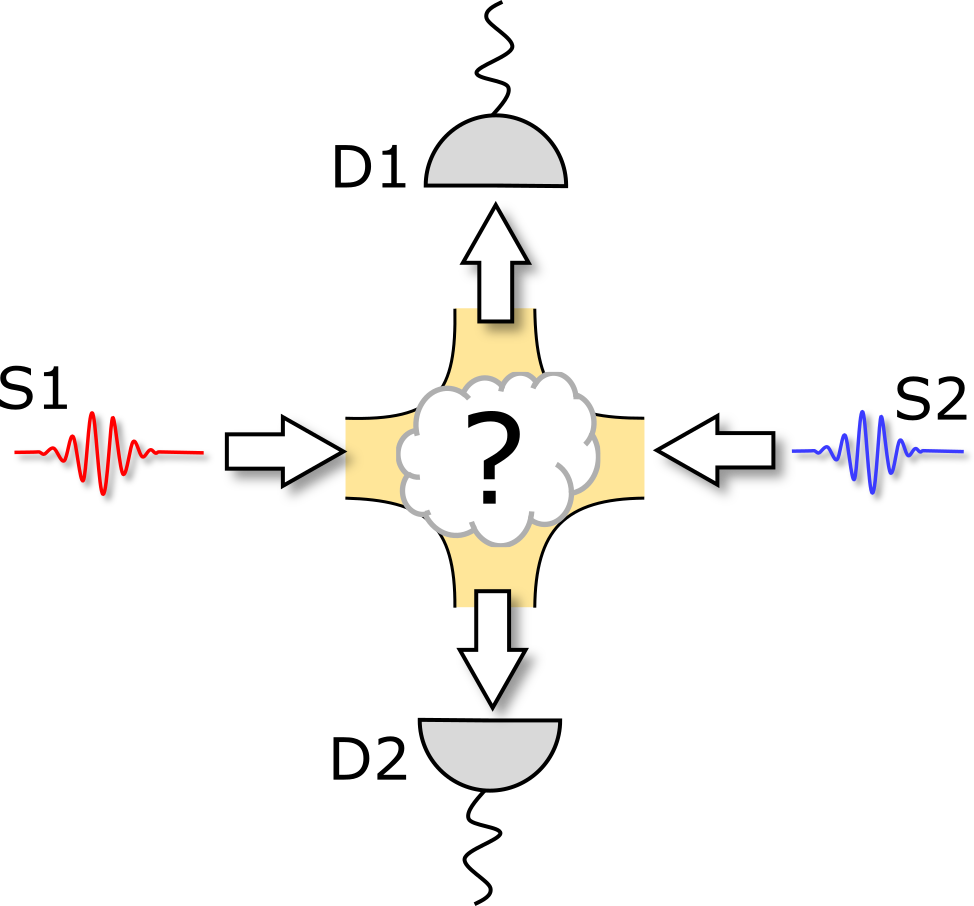}
\caption{%
{\bf Idealised electronic wavepacket collider} Sources S1, S2 inject electrons into a collision region with a variable time delay. They interact and scatter into detectors D1 and D2. This can be used to study interactions between electrons, interactions with the environment, and behaviour of the electron sources.
}
\label{fig:fig0}
\end{figure}

In principle, time-resolved electronic interactions can be studied with a wavepacket collider like that sketched in Fig.~\ref{fig:fig0}. Single electron sources S1 and S2 emit particles with relative delay $t_{\rm 21}$ into an experimentally-defined collision region. Interactions determine how particles are partitioned into detectors D1 and D2 for different injection time.\cite{BocquillonScience2013,Dubois2013} Under some conditions, fermionic exchange effects can create antibunching of wavepackets, detected via reduced current noise at the detectors.\cite{BocquillonScience2013} This effect can be used as a measurement of the indistinguishability of the wavepackets, an important figure of merit for quantum coherent transport.\cite{Ferraro-2014-2} However, in general, understanding the behaviour of the electrons in the interaction region is not straightforward if direct Coulomb effects are present in addition to exchange effects.\cite{BellentaniPRB2019}

For sources injecting electrons near the Fermi energy, the impact of the Coulomb interaction on the electron trajectory is diminished by screening.\cite{BocquillonScience2013} Changes to the single electron trajectories or velocity from direct Coulomb interactions between single electrons has not been seen, although coupling to nearby conducting channels has been detected via decoherence of the wavepackets.\cite{freulon2015hong} Where electronic density is reduced, or for particles injected at higher energy, the trajectories and velocity of electrons are expected to be significantly modified.\cite{ElinaArxiv2022,Sungguen2022} Understanding this regime is important for the controlled use of Coulomb interactions for quantum logic gates.\cite{Bellentani_PhysRevB.102.035417}

\begin{figure*}[!ht]
\includegraphics[width=0.9\textwidth]{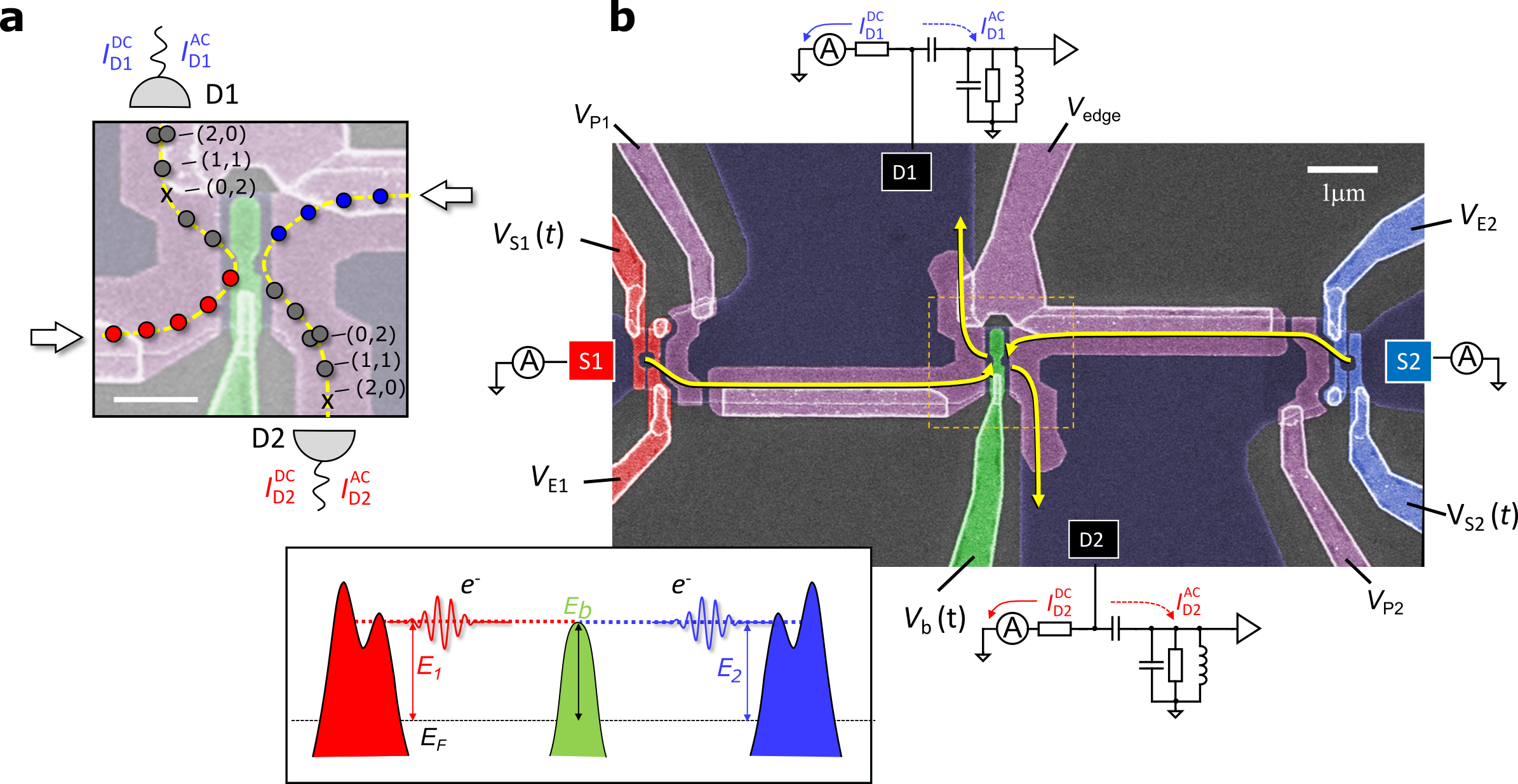}
\caption{%
\textbf{Electron collider implemented with high energy single electrons}
(a) Partitioning of electrons in a high energy device. Three outcomes are characterised by charge distributions (2,0) when both electrons enter D1, (0,2) when both enter D2, and (1,1) one enters each. Scalebar length is 500~nm. Proximity of electrons is exaggerated for clarity (see methods for timing information). 
(b) False colour scanning electron microscopy image of overall device structure and measurement system. See methods for details. Electrons emitted from sources S1 (left, red) and S2 (right, blue) eject electrons follow trajectories indicated by arrows, colliding at the central barrier (green). Dashed region indicates the collision region shown in (a). Inset: Energy level drawing to show adjustable ejection energy.
}
\label{fig:fig1}
\end{figure*}

To learn how to harness the Coulomb repulsion in ballistic electron systems, we have studied an electron collider based on electron pumps which emit electrons into edge states at high energy ($E> 100$~meV).\cite{fletcher2013clock,ubbelohde2015partitioning, waldie2015measurement,fletcher2019continuous} In this case, the time-resolved Coulomb interaction between ballistic single electrons can be directly detected.

\begin{figure*}[!ht]
\includegraphics[width=0.95\textwidth]{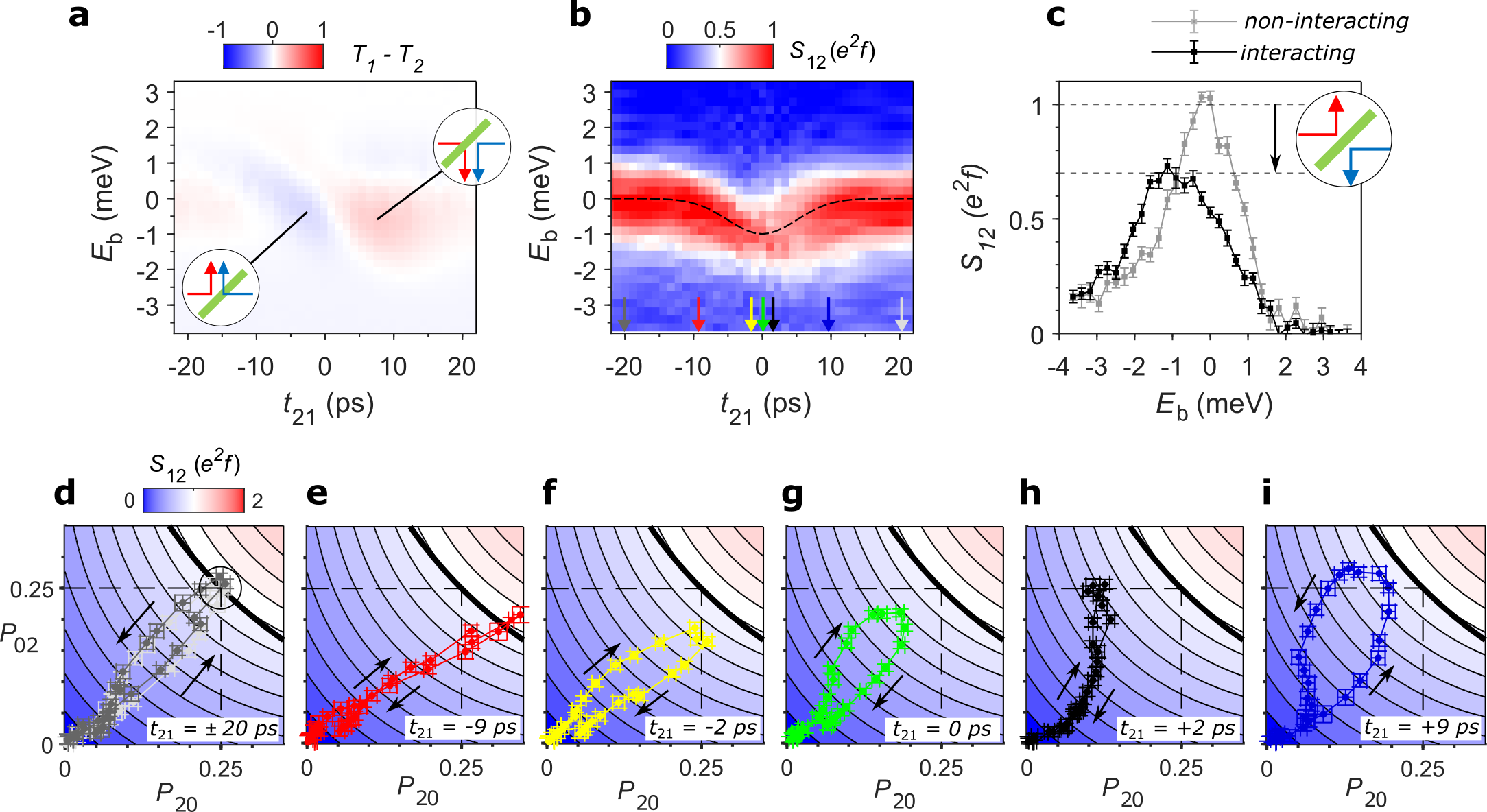}
\caption{%
\textbf{Current imbalance, shot noise suppression and partition statistics}
(a) Current imbalance $T_1 - T_2$ for different barrier height $E_{\rm b}$ and time delay $t_{\rm 21}$. Favoured events associated with the negative and positive current imbalances are indicated schematically. %
(b) Map of shot noise $S_{12}$ for barrier height $E_{\rm b}$ and time delay $t_{\rm 21}$. Dashed line is a guide to peak noise position. %
(c) Shot noise $S_{12}$ in the non-interacting ($t_{\rm 21} = -20$~ps) and strongly interacting regime ($t_{\rm 21} = +2$~ps) where the effect is most pronounced. Reduction in noise is associated with repulsion events as indicated. For estimation of errors bars, see Supplementary Section 1. %
Each panel (d-i) shows partition probabilities ($P_{20}$, $P_{02}$) at different barrier heights for a chosen $t_{21}$. Underlying colour map shows noise $S_{12}$ (contour lines are at $0.1e^2f$ intervals).
(d) The non-interacting case (electrons miss each other) at $t_{\rm 21} = -20$~ps (dark grey) +$20$~ps (light grey). Arrows show the sweep from high to low barrier height, reaching the maximum noise at independent scattering limit (thick solid line). At this point $(P_{20},P_{02}) = (0.25,0.25)$ (circled) there are an equal statistical mixture of each outcome event.  %
(e-i) as (d) but for selected interacting cases $t_{\rm 21} = -9, -2, 0, +2, +9$~ps corresponding to arrows in panel b. These cases show a strongly modified behaviour as described in the text.}
\label{fig:fig2}
\end{figure*}

The collision region of the electron collider is shown in Fig.~\ref{fig:fig1}a and the overall device structure in Fig.~\ref{fig:fig1}b. Electron pump sources S1 (left) and S2 (right) each inject an electron every $\tau = 2$~ns.\cite{giblin2012towards} In a perpendicular magnetic field these are confined to states on the mesa-edge with energies $E_1$ and $E_2$ typically in excess of 100~meV.\cite{fletcher2013clock} These can be individually tuned\cite{fletcher2013clock} such that $E_1 \simeq E_2$. Trajectories meet at a barrier with height $E_{\rm b}$ set by a gate voltage (see methods). At the barrier, electrons from source 1 and 2 are transmitted with probabilities $T_1$ and $T_2$ into output arms D2 and D1, respectively. The emission times from the two sources can be fine-tuned so that the nominal difference between arrival times at the centre barrier $t_{\rm 21} = t_2 - t_1$ can be chosen arbitrarily. For $t_{\rm 21}=0$ arrival is synchronised (see methods), down to a residual uncertainty determined by the emission time variation of the sources, typically of order picoseconds.\cite{fletcher2019continuous}

When only one source is active, individual $T_n(E_{\rm b})$ can be measured from the time-averaged current.\cite{fletcher2013clock,fletcher2019continuous} For our two sources we find that $T_1(E_{\rm b}) \simeq T_2(E_{\rm b})$. When both sources are active, the current at detector D2 is $I_{\rm D2}^{\rm DC} = ef(T_1 + 1-T_2)$ i.e. collecting any electrons from source S1 transmitted through the barrier and those from source S2 reflected by the barrier. Similarly $I_{\rm D1}^{\rm DC} = ef(T_2 + 1-T_1)$. 

For deliberately mismatched arrival times $I_{\rm D1}^{\rm DC} = I_{\rm D2}^{\rm DC}  \simeq ef$, a nearly constant background. When arrival times are synchronised to within a few picoseconds (Supplementary Section 1) an imbalance in transmission is seen. We show this, calculated using $T_1-T_2 = (I_{\rm D2}^{\rm DC} - I_{\rm D1}^{\rm DC})/2ef$, in Fig.~\ref{fig:fig2}a. For values near $t_{\rm 21} = 0$ we see a pronounced current imbalance with $T_1 - T_2 \simeq +0.2$ for $t_{\rm 21}>0$ and $T_1 - T_2 \simeq - 0.2$ for $t_{\rm 21}<0$. The polarity of this signal corresponds to preferential transmission of the earlier electron.

Measuring the time-dependent noise current, $I_{\rm D1}^{\rm AC}, I_{\rm D2}^{\rm AC}$ using the additional connections shown in Fig.~\ref{fig:fig1}b enables us to measure correlations\cite{BocquillonScience2013,ubbelohde2015partitioning} in partitioning outcomes from individual scattering cycles that are not accessible in a time-averaged measurement. The shot noise current $S_{12}$, simultaneously measured in contacts D1 and D2 (see methods), is mapped as a function of $E_{\rm b}$ and $t_{\rm 21}$ in Fig.~\ref{fig:fig2}b. In the case $|t_{\rm 21}| \gg 0$, electrons arrive at very different times and do not interact. The noise signal in this regime is explained by independent stochastic transmission of the electrons into either D1 or D2 given by $S_{12} = 2e^2 f\sum_{n=1,2} [T_n(1-T_n)]$.\cite{ubbelohde2015partitioning} The noise $S_{12} \simeq 0$ when the barrier is much higher or lower than the injection energy, but reaches a peak $S_{12}|_{\rm max} = e^2f$ when $E_{\rm b} \simeq E_1 \simeq E_2$ and $T_1, T_2 \simeq 0.5$ (i.e. the noise peak can pick out the electron energy).

The noise is modified in several ways by Coulomb interactions. Firstly, for values of $t_{\rm 21} \simeq 0$ (i.e. electrons collide), the position of the noise peak moves to a lower value of the programmed barrier height $E_{\rm b}$. This apparent shift in detected energy, $E_1 \rightarrow E_1'$ and $E_2 \rightarrow E_2'$ is a manifestation of the Coulomb repulsion between the electrons which gives a transient increase in the effective barrier height of $\simeq 0.5-1$~meV. Conservation of energy is preserved by a reduction in the kinetic energy corresponding to the increased Coulomb repulsion.\cite{Sungguen2022} Secondly, at delay times near $t_{\rm 21} = 0 $ the maximum partition noise $S_{12}|_{\rm max}$ is reduced from the independent scattering limit by up to $\simeq 30$\%, as shown in Fig.\ref{fig:fig2}(c). This effect is driven by an increase in the number of events where both electrons are reflected due to a repulsive interaction. While the noise reduction is in superficial resemblance to a partial Pauli dip,\cite{BocquillonScience2013} we show below that the full partitioning statistics, calculated from the time-averaged current and noise data,\cite{ubbelohde2015partitioning} reveal a Coulomb-dominated system.

The partition probabilities for the three possible detector charge states $P_{ij}$ are $P_{20}, P_{11}, P_{02}$ where $i,j$ are the number of electrons in detector D1, D2 and $\sum P_{ij} = 1 $. Particular values of $P_{02}, P_{20}$ map directly to current noise and time-averaged current distribution.\cite{ubbelohde2015partitioning,Hassler} Computed values of $P_{02}, P_{20}$ at different barrier height are shown in Fig.~\ref{fig:fig2}d-i. The non-interacting case in Fig.~\ref{fig:fig2}d ($|t_{\rm 21}| = 20$~ps) has $P_{02}, P_{20} \simeq 0, P_{11} \simeq 1$ for high and low barriers and the expected statistical mixture of outcomes when $E_{\rm b} \simeq E_{1}, E_{2}$ namely $P_{02}|_{\rm max} , P_{20}|_{\rm max} \simeq 0.25$ and $P_{11}|_{\rm min} \simeq 0.5$. The results in the interacting regime ($|t_{\rm 21}| < 10$~ps) are shown in Fig.~\ref{fig:fig2}e-i. Preferential transmission into D1 or D2 causes an imbalance in $P_{20}$ and $P_{02}$ (a tilt toward the $P_{02}$ or $P_{20}$ axis). $P_{20}|_{\rm max}$ and $P_{02}|_{\rm max}$ are also suppressed below the independent scattering value $\sqrt{P_{20}} + \sqrt{P_{02}} = 1$ (thick solid line in Fig.~\ref{fig:fig2}d-i). This corresponds to a reduction in noise or an excess antibunching $\Delta P_{11} = P_{11}|_{\rm min} - 0.5 \simeq 0.15$. 

We show below that the full partitioning dataset is fully captured by a model of the microscopic trajectories which follow the $E \times B$ drift motion. The shape of these trajectories is set by the partitioning barrier potential and the (weakly screened) Coulomb interaction. We also show that the small timing and energy fluctuations of the source\cite{fletcher2019continuous} are important in determining the statistical outcomes seen experimentally.

\begin{figure*}[t]
\includegraphics[width=0.85\textwidth]{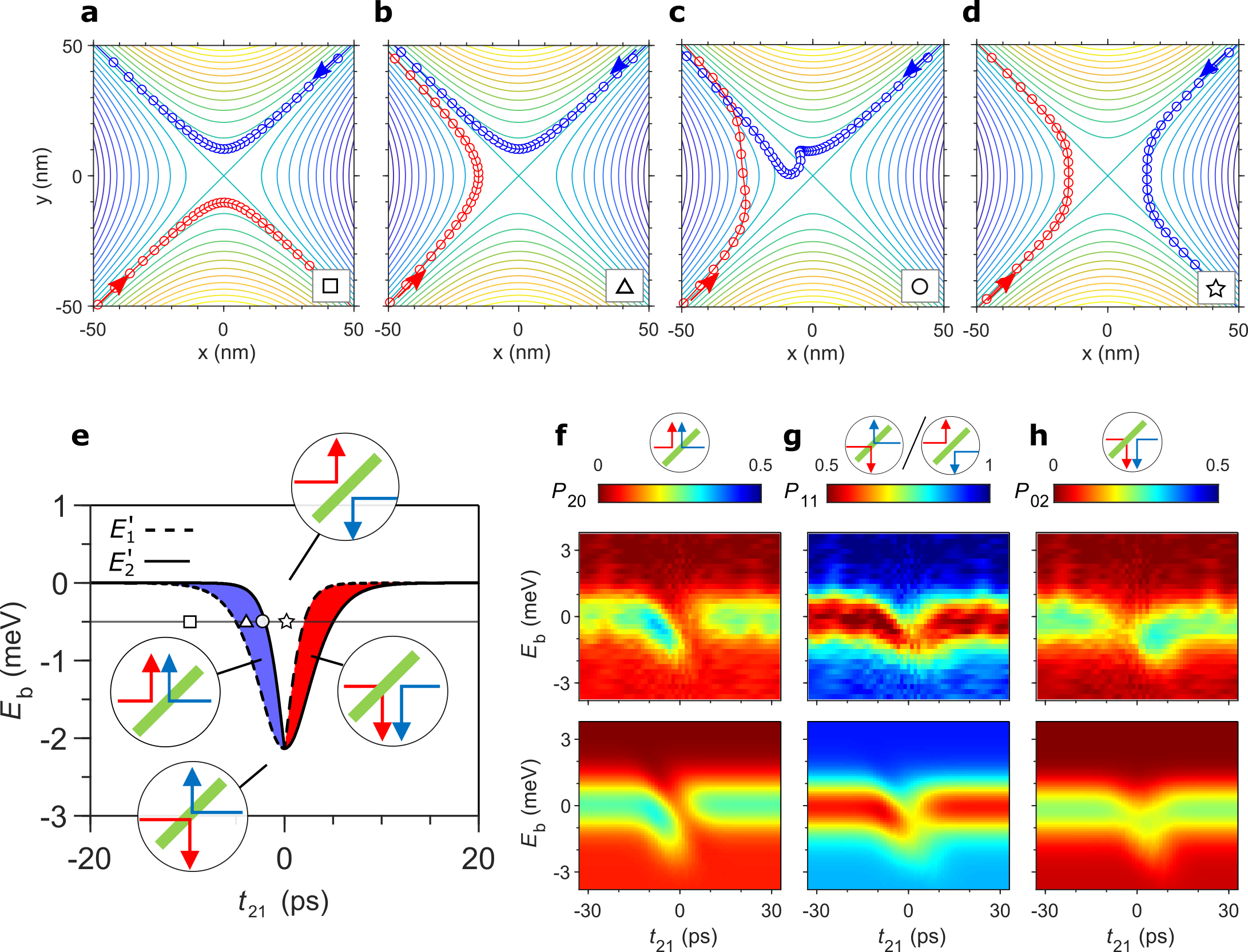}
\caption{%
\textbf{Trajectory of interacting electrons and the resulting partition statistics:} 
(a-d) Electron trajectories for electrons colliding near the saddle point barrier at $E_1 = E_2 = E_{\rm b} + 0.5$~meV. (a)~$t_{\rm 21} = -10$~ps non-interacting case; the electron from source S2 arrives first. Electrons from source S1 and S2 are both transmitted. (b) $t_{\rm 21} = -4$~ps; S2 electron arrives first and is transmitted, but blocks S1 electron. %
(c) $t_{\rm 21} = -2.5$~ps; same outcome as (b), but exit of the S2 electron from the scattering region is delayed. (d) $t_{\rm 21} = 0$~ps; electrons block each other at the barrier and both are reflected. By symmetry, corresponding trajectories are found for $t_{\rm 21}>0$ (not shown). %
(e) Transmission imbalance $T_1 - T_2= -1,0,1$ (blue, white, red) as a function of $t_{\rm 21}$ and $E_{\rm b}$ for $E_1 \simeq E_2$. Four regions of uniform $T_1 - T_2$ are seen. Two have uneven transmission distribution (areas shaded red and blue) while two have an even charge distribution (shaded white) which correspond to cases where electrons are are both reflected or both transmitted. Dotted ($E_1'$) and solid ($E_2'$) lines mark individual transmission thresholds for electrons from source S1 and S2. Symbols ($\square, \triangle, \medcircle, \medstar$) indicate the specific values of $(t_{\rm 21},E_{\rm b})$ used for the trajectories in panels a-d. %
(f), (g), (h) upper panels show full maps of experimentally obtained probabilities $P_{20}$,  $P_{11}$, $P_{02}$. Lower panels show calculation of ensemble average over classical outcomes
}
\label{fig:fig3}
\end{figure*}

We compute the trajectories for an assumed form of the potential\cite{fertig1987transmission} near the barrier $U_{\rm 2D}(x,y)$ and an interaction potential $U_{ee}(\vec{r_1},\vec{r_2}) \propto 1/r$ at short range, but which weakens at longer distances ($r_{12} \gg 90$~nm, see methods) due to screening by metallic surface gates.\cite{SkinnerPhysRevB.82.155111} Interactions are limited to a region within the screening length near the effective tunneling point, making a saddle point barrier potential an appropriate approximation.\cite{fertig1987transmission} Calculations based on the $E \times B$ trajectories are valid in the regime of our experiments where the energy window of quantum scattering (over which the barrier transmission probability changes from 0 to 1) is much smaller than the energy uncertainty of the electrons and/or the purity of the electrons is sufficiently low \cite{fletcher2019continuous} (see Supplementary Section 2).

Example trajectories are shown in Fig.~\ref{fig:fig3} for injection energy exceeding the barrier height $E_n  - E_{\rm b} = 0.5$~meV. This illustrates three important cases; completely mismatched arrival Fig.~\ref{fig:fig3}a, near-synchronised Fig.~\ref{fig:fig3}b,c and closely synchronised Fig.~\ref{fig:fig3}d. In the non-interacting case Fig.~\ref{fig:fig3}a, trajectories follow the equipotential contours of $U_{2D}$ and the transmitted charge $(Q_{T1},Q_{T2}) = (1e,1e)$. In contrast, close to synchronisation as in Fig.~\ref{fig:fig3}d, both electrons are deflected such that $(Q_{T1},Q_{T2}) = (0e,0e)$. For the near-synchronised cases in Fig.~\ref{fig:fig3}b,c interactions modify the trajectory of the electrons in a more complicated way. The late-arriving electron (from S1 in these examples) tends to be deflected by the earlier-arriving electron (here from S2) which blocks the barrier region. This leads to $(Q_{T1},Q_{T2}) = (0e,1e)$ for a range of $t_{\rm 21}$ including examples like those shown in Fig.~\ref{fig:fig3}b and c. 

Establishing that the order-of-arrival effect described above is accessible experimentally would show that the Coulomb interaction can be controlled on the time-scales relevant for interactions in edge states. We numerically compute a large number of trajectories at different injection energies $E_1-E_b, E_2-E_b$ and relative arrival time $t_{21}$, making a detailed map of the charge partitioning outcomes (see Supplementary Section 2). We use this to reveal how details of the Coulomb interaction appear in experimental partitioning data.

The transmission imbalance $T_1 - T_2$ is plotted in Fig.~\ref{fig:fig3}e, computed from the charge distribution $(Q_{T1}, Q_{T2})$ at various arrival time delays and barrier heights for equal injection energy $E_1 = E_2$. This map consists of four regions, encompassing the specific example trajectories in Fig.~\ref{fig:fig3}a-d (see markers). Each region is bounded by the values of $E_{\rm b} = E_n'$, the programmed barrier height at which the first or second electrons stop being transmitted in the presence of Coulomb repulsion. The functional form of $E_{n}'(t_{\rm 21})$ is an asymmetric peak created by the accumulated effect of the Coulomb interaction during the interaction. In the region where the peaks overlap, there is a regime where the electrons repel, as in Fig.\ref{fig:fig3}(d). At larger values of $|t_{\rm 21}|$, when electron arrival is not exactly synchronised, the asymmetry gives $E_1'(t_{\rm 21}) \neq E_2'(t_{\rm 21})$. This is because generally the effective kinetic energy loss of each particle is different. This creates an intermediate region where only the late-arriving electron is blocked. This predicts a current imbalance in $T_1 - T_2 = -1, 0, +1$ with the same polarity as that measured experimentally in Fig.~\ref{fig:fig2}a. This suggests that the fine structure in the time-resolved Coulomb collision described above is indeed visible experimentally, but appears broadened.

In this calculation, we find that the exact values of $E_n'$ and the range of $t_{\rm 21}$ over which the interaction extends $t_\omega$ are governed by the cyclotron frequency, the curvature of the saddle point $\omega_{xy} = \sqrt{\omega_x \omega_y}$ and the inclusion of screening. However, the qualitative behaviour of the partitioning map is not sensitive to the precise parameters used. A more important consideration for comparison with experiment is that the map of outcomes in Fig.~\ref{fig:fig3}e captures classical behaviour for well-defined injection energies and times. A finite phase-space area for electronic injection energy and time is expected\cite{Kashcheyevs2017} along with extrinsic emission time and energy fluctuations of the source.\cite{fletcher2019continuous} These effects will broaden the boundaries of the different scattering outcomes as the barrier height and time of arrival vary. This has an impact on the maximum visibility of the order-of-arrival effects, completely smearing them out in the case of grossly time-broadened emission (see Supplementary Section 1). 


We compute the partition probabilities $P_{20}, P_{11}, P_{02}$ from the ensemble average of the charge distribution $(Q_{T1},Q_{T2})$ over the initial conditions $(E_1,E_2,t_{21})$ of the trajectories obtained from the source energy-time distribution (i.e. energy, time widths $\sigma_t$ and $\sigma_E$ measured for each source, see methods). 

The full experimental partition statistics and the computed ensemble average are shown together in Fig~\ref{fig:fig3}f-h. The agreement between the experimental data (upper panels) and model calculations (lower panels) show that the Coulomb interaction, broadened by the fluctuations of the electrons sources, is accessible experimentally. Overall, the ensemble effects blurs the sharp boundaries of the classical outcomes, reducing the size of the shift in the noise peak. The energy-time correlation present in the emission distributions, a characteristic feature of the source,\cite{fletcher2019continuous} and mismatches between the exact source emission distributions explains the fine structure of the partitioning data. That the energy shift and the order-of-arrival effect are visible shows that that the sharpness of our emission distributions are on the same picosecond time scale required to control single electron interactions with high fidelity.

Single-electron antibunching signatures driven by the Pauli effect were observed with `mesoscopic capacitor' sources which alternately emit electrons and holes.\cite{Bocquillon-2014-1} In that case, as expected of fermionic identical-particle exchange statistics, noise suppression was seen for electron-electron synchronisation but not for electron-hole synchronisation.\cite{BocquillonScience2013} This was an important observation to rule out the possible roles of Coulomb effects. We have to consider here the reverse question: In a system with relatively strong Coulomb interactions, is it ever possible for fermionic exchange effects to play a role? The technical factors that might prevent this include the limited purity of the sources\cite{fletcher2019continuous} and the energy selectivity of the barrier, which modifies the electronic wavepacket.\cite{BellentaniPRB2019,Sungguen2022} More fundamentally, it appears from our measurements and a realistic model that the trajectories of the electrons are so strongly repelled that achieving a high overlap of the incoming wavepackets is unlikely, at least in this parameter regime, and there would be no manifestation of fermionic exchange effects even for perfect sources and an energy-independent barrier.

As far as we have seen, Coulomb effects are visible provided the fluctuations in arrival difference $|t_2 - t_1|$ of the sources are not too much larger than the interaction time scale $t_\omega$ of the barrier (in the present devices $\sigma_{t1},\sigma_{t2} \geq t_{\omega}$). This technique actually provides a very sensitive readout of the emission-time distribution of single electron sources without the use of high bandwidth probe signals.\cite{fletcher2019continuous} This is somewhat in the spirit of the original Hong Ou Mandel\cite{Hong-1987-1} experiment in optics which first used two particle interference to measure the properties of laser sources on sub-picosecond time scales. Particle colliders, as considered in Figure~\ref{fig:fig0}, can provide information, regardless of the nature of the interaction. Here this approach can be used to optimise the emission characteristics of on-demand sources\cite{fletcher2019continuous} for applications in time-resolved sensing\cite{Johnson2017} and other high speed applications.

In summary we have revealed the effect of mutual repulsion between single electrons in high-energy chiral ballistic channels using an electron collider. The partition statistics which probe indistinguishability in highly screened regimes instead reveal here Coulomb-dominated perturbations of electron trajectories in a depleted channel at high magnetic field. It will be interesting to further trace the perturbation of the trajectories by performing tomographic measurements.\cite{fletcher2019continuous} At much lower energy, the interplay of the fermionic exchange statistics and the Coulomb interactions may be accessible in a weak interaction regime achieved by electrostatic gate arrangements.\cite{Duprez-2019-1}

We have recently learned of experiments performed with electrons sources similar to this, but operating slightly closer to the Fermi Energy (30-60 meV) using single shot charge detection.\cite{UbbelohdeChargeDet} These experiments also suggest a Coulomb-dominated regime that extends over a broad range of energy. Experiments with electrons confined in travelling surface acoustic wave minima\cite{WangSAW} also shown an apparent antibunching that may be driven by Coulomb repulsion. This shows that the Coulomb interaction may dominate in a range of scenarios relevant for electron quantum optics or flying qubits.

\bibliography{Collision.bib} 
\bibliographystyle{apsrev}

\newpage

\section{Acknowledgements}

We acknowledge experimental assistance from Nathan Johnson and Shota Norimoto. 
W.P. and H.-S.S. acknowledge support by Korea NRF via the SRC Center for Quantum Coherence in Condensed Matter (Grant No. 2016R1A5A1008184).
This work was supported by the UK government's Department for Business, Energy and Industrial Strategy and from the Joint Research Projects 17FUN04 SEQUOIA from the European Metrology Programme for Innovation and Research (EMPIR) cofinanced by the Participating States and from the European Unions Horizon 2020 research and innovation programme. SR acknowledges support from the Maria de Maeztu Program for Units of Excellence No. MDM2017-0711 funded by MCIN/AEI/10.13039/501100011033.

\section{Methods}

\textit{Device:} Our sources are fabricated on GaAs/AlGaAs two dimensional electron gas (2DEG) wafer depth $d_{\rm 2DEG} = 90$~nm, carrier density $n_{\rm 2d} = 2.4\times 10^{11}$cm$^{-2}$ mobility $\mu = 2.5 \times 10^6$cm$^2$/(V.s). Each pump is a gate-defined quantum dot in a channel 800~nm wide at a distance of 5~$\mu$m from the centre barrier. Entrance gates, adjacent to the source reservoir, are controlled by $V_{\rm S1,S2}(t)$, modulated to load electrons from the source reservoirs. The same modulation pushes the trapped electron towards the exit barriers controlled by $V_{\rm E1,E2}(t)$~\cite{Giblin2020,fletcher2013clock,Leicht_2011}. Interaction effects have been explored with both sinusoidal and tailored pump waveforms at several frequencies. Data presented here is from a sinusoidal drive at $f = 500$~MHz. The device is measured in a perpendicular magnetic field $B = 10$~T (into the plane of Figure~\ref{fig:fig1}) in a dry dilution refrigerator with mixing chamber $T_{\rm mc}  < 0.1$~K. 

\textit{Energy tuning and measurement:}
The pump exit barrier controls the number of electrons collected as well as the energy of the injected electrons.\cite{fletcher2013clock} This is used to adjust/match the injection energies of the two sources. Relative electron energy can be measured from the setting of $E_{\rm b}$ required to block transmission. Gate voltage values scaled to energy using the phonon emission features in the energy distribution at multiples of $\hbar \omega_{\rm LO} = 36$ meV.\cite{Johnson2017} The absolute injection energy requires a separate calibration experiments not performed here, but the presence of $n_{\rm LO} = 3$ side-bands to the main electron distribution at a spacing of $\hbar \omega_{\rm LO}$ indicates an energy of at least $\sim 100~$meV as in previous reports.\cite{fletcher2013clock}

\textit{Edge state transport:}  
In a perpendicular magnetic field, $B = 10~T$, the electrons emitted from the pump follows a chiral edge channel towards the centre of the device. Electrons follow high energy Landau levels with drift velocity $v_d = (\vec{E} \times \vec{B})/B^2$ governed by the edge electric field $\vec{E}$ and magnetic field.\cite{Kataoka2016}. Relaxation to lower energy states is strongly inhibited by the edge gates\cite{Johnson2018} here controlled by $V_{\rm P1}$, $V_{\rm P2}$, $V_{\rm edge}$ as described in the Supplementary. A small residual phonon emission means that $\sim 5\%$ of the injected electron are always reflected. This creates a small residual noise when $E_n-\hbar \omega_{\rm LO} >E_b<E_n$, visible as a residual value of $P_{20},P_{02}>0$ in Figure~\ref{fig:fig2}d-i for low barriers. This effect is included in the computed partition statistics in Figure~\ref{fig:fig3}.

\textit{Time-of-arrival synchronisation:} 
To find the delay setting corresponding to $t_{12} = 0 $ a synchronisation pulse is applied to the centre barrier synchronised with both electron pumps but with an adjustable delay $t_b$. Maps of drain current versus $E_{\rm b}$ and $t_b$ reveal the shape of the synchronisation pulse, referenced to each electron's arrival time (see Ref.\cite{Johnson2017} and Supplementary). Note that the transit time estmated from the velocity\cite{Kataoka2016} $t_{\rm path} \leq 100~ps$ is shorter than the repeat period $1/f_{p} = $2000~ps so there are no interactions between electrons emitted in different cycles.

\textit{Partition readout:} Average current in output channels $I_{D1}$ and $I_{D2}$ is read with current-to-voltage converters. Using $I_{\rm D2} + I_{\rm D2} = 2ef$ the data in Fig.3a is calculated from $T_1-T_2 = (I_{\rm D2}^{\rm DC} - I_{\rm D1}^{\rm DC})/2ef = I_{\rm D2}/ef - 1$. Cross-correlated shot noise $S_{12}$ is also measured on D1 and D2, calibrated using single electron partitioning. Here we consider the absolute value only; the measured value of $S_{12}$ is negative due to the detector configuration.\cite{ubbelohde2015partitioning}  See Supplementary Section 1 for further details.

\textit{Calculation of trajectories:}
The trajectories of electrons in a 2D plane are obtained based with a total potential $U_{\rm tot} (\vec{r_1},\vec{r_2}) = U_{\rm 2D}(\vec{r_1}) + U_{\rm 2D}(\vec{r_2}) + U_{ee}(\vec{r_1},\vec{r_2})$ with $U_{\rm 2D} = E_{\rm b} - m^*\omega_x^2x^2/2 + m^*\omega_y^2y^2/2$. Numerical integration of the equations of motion gives trajectories and the final charge distribution can be evaluated after electrons leave the scattering area (see Supplementary Section 2). These calculations are similar to those of Ref.\cite{ElinaArxiv2022}. We define $U_{ee}$ to include a direct contribution from the interaction between electrons in the GaAs material ($\epsilon_r = 12.9$) and an interaction with an image charge at a distance $2d_{\rm 2DEG}$ from the plane\cite{SkinnerPhysRevB.82.155111} which screens the interaction at distances $r_{12} \gg d_{\rm 2DEG} = 90$~nm. This weakens the interaction and makes the effects of the potential away from the scattering region less significant. Selection of initial particle positions sets both electronic energy and the relative injection time (this uses the analytical solutions to the non-interacting equations of motion appropriate for large $r_{12}$). See Supplementary Section 2. We choose a single value of $\omega_x = \omega_y = 5 \times 10^{12}$ s$^{-1}$ to reproduce the experimentally observed features after accounting for the effect of the input energy-time distributions (see below). The qualitative features of the scattering outcome map (the regions labelled in Fig.~\ref{fig:fig3}e do not change significantly in this 2D regime with exact choice of $\omega_x, \omega_y$.\cite{ElinaArxiv2022} The temporal width of the dip in $E_1'$ and $E_2'$ become narrower with increasing $\omega_x, \omega_y$ as the electrons are forced closer together.

\textit{Ensemble average over input energy, time distributions:} 

To simulate the effect of stochastic fluctuations in the initial energy-time distribution we first compute classical outcomes for discrete values of  different input energies $E_1, E_2$ and times $t_1, t_2$ in a high resolution grid, then sample these outcomes using distributions based on experimentally observed input distributions for the two sources $W_1(E_1,t_1)$ and $W_2(E_2,t_2)$. See Supplementary Section 2. The partition noise, current imbalance are readily computed from the sampled outcomes, from which the partition statistics can be calculated. Rather than a full tomography of the incoming Wigner distributions~\cite{fletcher2019continuous} we use a simpler technique which parameterises the Wigner distribution as a bivariate gaussian $W_n(\sigma_t, \sigma_E, r)$ with energy and time widths $\sigma_E$ and $\sigma_t$ and energy time correlation coefficient $r$.

\begin{equation}
    W_n = \frac{1}{2\pi \sigma_t \sigma_E \sqrt{1-r^2}} 
    \exp{ 
        \left[ -\frac{1}{2(1-r^2)} \left( \frac{E^2}{\sigma_E^2} + \frac{t^2}{\sigma_t^2} - \frac{2rEt}{\sigma_E \sigma_t} \right)
        \right]}
\end{equation}

We find the size of the emission distributions similar to that found previously\cite{fletcher2019continuous} with the most significant difference being the temporal length of the two electrons is different $\sigma_{t1} = $1.7 $\pm$ 0.5~ps $\sigma_{t2} = $5.2 $\pm$ 1~ps while the injected energy distributions are similar, $\sigma_{E1} = 0.85 \pm 0.2$ ~meV and $\sigma_{E2} = 1.05 \pm 0.2$~meV (See Supplementary Section 1). We compute the partition statistics over $N = 10^4$ samples from distributions with these parameters over the maps of scattering outcomes.

\section{Author contributions}

JDF designed and developed measurement system and experimental methodology. With input from PS and MK, designed samples. Performed experiment, data acquisition and analysis. With WP, helped to explore numerical calculation of particle trajectories and comparison with data. 

WP developed a method of calculation of classical particle trajectories. Performed numerical calculations to compare with experimental data. Investigate methods to establish the validity of the classical model. 

SR developed 1D quantum collision model (published separately) and helped to develop a classical method of calculation of particle trajectories and advise on its realm of validity. 

PS developed fabrication techniques for electron pump devices. Fabricated samples used in this paper and preliminary batches of prototype electron colliders. 

JPG, GACJ provided electron beam sample patterning. 

IF, DAR provided wafers for the device substrates. 

HSS oversaw the development of both quantum and classical models of particle collision. 

MK directed this research project. Supported JDF for experiments and data analysis. Suggested the underpinning mechanisms that result in positive and negative correlation in two electron transmission/reflection observed experimentally. 

The manuscript was written by JDF, MK, WP, SR and HSS with review by other authors.

\end{document}


\title{Supplementary Information for `Time-resolved collision of single electrons'}

\author{J.D. Fletcher}
\email{jonathan.fletcher@npl.co.uk}
\affiliation{National Physical Laboratory, Hampton Road, Teddington TW11 0LW, United Kingdom}
\author{W. Park} 
\affiliation{Department of Physics, Korea Advanced Institute of Science and Technology, Daejeon 34141, Korea}
\author{S. Ryu} 
\affiliation{Instituto de F\'{i}sica Interdisciplinary Sistemas Complejos IFISC (CSIC-UIB), E-07122 Palma de Mallorca, Spain}
\author{P. See} 
\affiliation{National Physical Laboratory, Hampton Road, Teddington TW11 0LW, United Kingdom}
\author{J.P. Griffiths} 
\affiliation{Cavendish Laboratory, University of Cambridge, J. J. Thomson Avenue, Cambridge CB3 0HE, United Kingdom}
\author{G.A.C. Jones} 
\affiliation{Cavendish Laboratory, University of Cambridge, J. J. Thomson Avenue, Cambridge CB3 0HE, United Kingdom}
\author{I. Farrer} 
\affiliation{Cavendish Laboratory, University of Cambridge, J. J. Thomson Avenue, Cambridge CB3 0HE, United Kingdom}
\author{D.A. Ritchie} 
\affiliation{Cavendish Laboratory, University of Cambridge, J. J. Thomson Avenue, Cambridge CB3 0HE, United Kingdom}
\author{H.-S. Sim}
\affiliation{Department of Physics, Korea Advanced Institute of Science and Technology, Daejeon 34141, Korea}
\author{M. Kataoka}
\affiliation{National Physical Laboratory, Hampton Road, Teddington TW11 0LW, United Kingdom}

\date{\today} 
\begin{abstract}
\end{abstract}
\maketitle

\section{Experimental Details}

\subsection{Control signals}

We use three synchronised Tektronix 70001B Arbitrary Waveform Generators (AWG) to create pump drive and timing radio frequency (RF) signals. These instruments have a sample rate of 50 GS/s, an analog bandwidth of 15~GHz and programmable channel-to-channel phase resolution of $\simeq 0.22$~ps. After broadband amplification (+15 dB) and attenuation/losses in the coaxial lines the approximate amplitude at the devices is $\sim 200-400$~mVpp. This amplitude is adjusted to cross electron loading/ejection thresholds.\cite{giblin2012towards} For the data shown in the main paper the pump drive signals are sine waves with $f = 500$~MHz. These are filtered with an 8 GHz low pass filter to remove residual leakage from the AWG master clock ($12.5$~GHz) which might otherwise disturb the monotonicity of the drive waveform near emission.\cite{waldie2015measurement} DC voltage sources are low-pass filtered with $f_{3dB} < 10~$kHz and, where required, combined with RF signals using a resistor/capacitor bias tee at low temperature.

\subsection{Noise and DC current Readout}

The time average (DC) current $I_{\rm D1}^{\rm DC}, I_{\rm D2}^{\rm DC}$ in the two drain contacts is measured with commercial current-to-voltage converters (Femto DDPCA-300, $10^{10}$ V/A range). To measure shot noise currents $I_{\rm D1}^{\rm AC}, I_{\rm D2}^{\rm AC}$ we couple an amplification chain to each drain contact. Resistor-capacitor networks are used to isolate AC and DC branches. Frequency-matched LCR tank circuits ($L = 66 \mu $H, $C_p \simeq$  120~pF, $R_{\rm eff} \simeq 9$~k$\Omega$) define a pass-band centred $f_{0} \simeq 1.8$~MHz with bandwidth $\delta f = 150$~kHz. After amplification at $T = $3.5~K and at room temperature, both signals are sampled synchronously with digitising card (National Instruments model NI-5922, sample rate 10~MS/s) followed by real-time FFT analysis in $10^4$ point segments (i.e. 1~ms of sampling).\cite{dicarlo2006system} These autocorrelation and cross-correlation spectra are averaged and, ultimately, integrated over a frequency range that encompasses the pass band of the tank circuit filter.

Experimentally, at each value of $t_{21}$ we repeatedly measure the noise at different values of the energy barrier $E_b$. This cycle is repeated for long enough (typically 10 mins per point) to reach sufficient signal-to-noise ratio. Statistical errors can be estimated from the standard error in these repeated measurements. At $f = 500$~MHz each pump generates a maximum noise of $S_{\rm max} = 0.5 e^2 f = 6.4 \times 10^{-30}$A$^2/$Hz. The partition noise\cite{BocquillonScience2013} can be extracted either from the autocorrelation signals or the cross-correlation signals. Auto-correlation signals have a Johnson noise contribution from the tank circuit resistor.\cite{dicarlo2006system} This can be subtracted using an interleaved reference measurement to reliably sample this thermal background. The cross-correlation signal has the advantages of not having a large, time-dependent background to subtract and from averaging over two amplifiers.\cite{dicarlo2006system}

\subsection{Noise Calibration}

We have confirmed that the single-electron partitioning noise of individual pumps matches the expected shot noise signal of $S_{12} = -0.5e^2f$, as seen by others.\cite{ubbelohde2015partitioning} The negative sign is a consequence of measuring on both sides of the partition barrier but only the magnitude of the cross-correlated noise is important here. The total effective gain depends on the effective resistance of the tank circuits, the pass band of individual tank circuits, the frequency dependent gain of all amplifiers and the sampling card, and the exact integration band used. In a previous run, we have used the temperature dependent Johnson noise of the shunt resistor $R(T)$ for calibration of the pass-band and amplifier gains.\cite{dicarlo2006system} Here, to account for all factors in a single step we determine a single calibration factor in-situ from the partitioning of electrons from pumps operated in isolation. Partition noise $S_n$ for pumps $n = 1,2$ both follow $S_n  = \alpha T_n(1-T_n)$ where $T_n$ is taken from the transmitted DC current. From the random uncertainty in the calibration data and the small difference between results from the two pumps we estimate the uncertainty that $\Delta \alpha/ \alpha \leq 6\%$.

\subsection{Pump settings, phonon emission and energy barrier calibration}


The device tuning is similar to previous work.\cite{fletcher2013clock,Johnson2017,fletcher2019continuous,Kataoka2016} Electron emission and transport is characterised by the number of electrons pumped, the emission energy-time distribution, and other factors like the probability of inelastic relaxation via phonon emission. The number of electrons loaded into the pump is largely set by the time-independent exit barrier. This is set to result in one electron per cycle, established from the quantisation of the pumped current.\cite{giblin2012towards} The same barrier also tunes the emission energy of the pumps, used to match the emission energy of the two pumps.\cite{fletcher2013clock} Each pump has a time-varying barrier $V_{\rm S1}^{\rm AC}(t),V_{\rm S2}^{\rm AC}(t)$ whose voltage is modulated around a DC value $V_{\rm S1}^{\rm DC}$, $V_{\rm S1}^{\rm DC} \simeq -0.5$~V to enable electron loading and ejection.\cite{giblin2012towards} As $V_{\rm S1}^{\rm AC}$, $V_{\rm S2}^{\rm AC}$ directly controls the emission ejection time,\cite{waldie2015measurement} the phase of these signals can be shifted to control arrival time at the partitioning barrier.

As in previous work\cite{Kataoka2016} we use an edge gate to deplete the entire edge region. This is kept at a constant negative voltage $V_{\rm edge}$ = -0.32~V for the duration of the experiment. This strongly suppresses emission of optical phonons.\cite{Johnson2017,Kataoka2016} Additional gates, controlled by signals $V_{\rm P1}$ and $V_{\rm P2}$, just outside the pumps are used to create a similar depleted region. These gates slightly change the electron emission energy and timing and are fixed to an intermediate value different from $V_{\rm edge}$ to reduce the phonon emission while not excessively disturbing pumping. Here $V_{\rm P1} = - 0.10$~V and $V_{\rm P2} = - 0.28$~V. The weak residual signal from the phonon emission line enables use to calibrate changes in the relative barrier height $E_b$ in energy units using $\hbar \omega_{\rm LO} = 36.0$~meV.\cite{fletcher2019continuous,fletcher2013clock} Here $\delta V_{\rm LO}/\hbar \omega_{\rm LO} \simeq 2.2$~mV/meV. 

\subsection{Energy matching}

\begin{figure}[h]
	\centering
	\includegraphics[width = 0.35\linewidth]{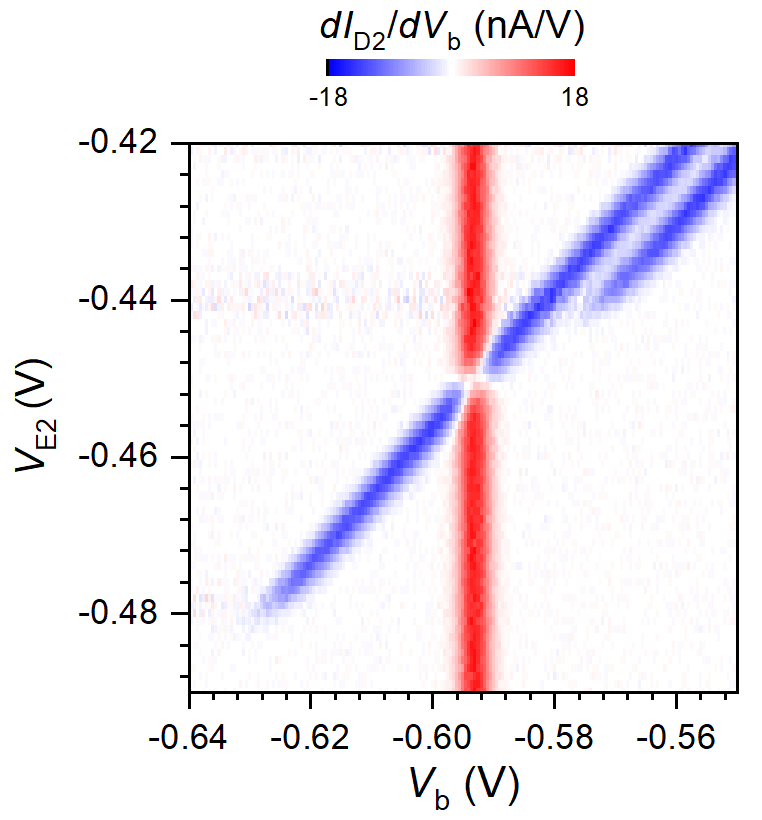}
	\caption{$dI_{\rm D2}/dV_{\rm b}$ with both pumps operating for different $V_{b}$ and $V_{\rm E2}$. This reveals sharp changes in current with barrier height as electrons from source S1 and S2 are blocked (note that the resulting changes in $I_{\rm D2}^{\rm DC}$ have opposite polarity). Here $V_{\rm E1} = -0.54$~V is constant, giving a constant positive peak in $dI_{\rm D2}/dV_{\rm b}$ (red line). $V_{\rm E2}$ controls the emission energy of source S2, detected as a negative peak in $dI_{\rm D2}/dV_{\rm b}$ (blue line). This enables a matching point in energy $E_1 = E_2$ to be found.}
	\label{fig:energy_matching}
\end{figure}

The exit barrier of each pump is the primary control over the number of electrons pumped, $n_{e1}, n_{e2}$. It also sets the energy of the injected electrons with a linear relationship $\Delta E_1 \propto \Delta V_{E1}$.\cite{fletcher2013clock}  We use this to match the injection energies of the two sources. Relative electron energy is measured from the setting of $E_{\rm b}$ required to block transmission, which is apparent from a reduction or increase in current into either D1 or D2. An example is shown in Fig.~\ref{fig:energy_matching}.

The width of the $n_{e} = 1$ pump current plateau sets an energy-tuning window. The position of the window can be moved somewhat using other parameters (e.g. the phonon-emission gates described above). In practise matching the number of loaded electrons $n_{e1} = n_{e2} = 1$ while simultaneously matching energy $E_1 = E_2$ is not always guaranteed. Even lithographically identical devices have slightly different loading and ejection behaviour and need to be tuned slightly differently.  Nonetheless, we have found overlapping emission energy windows in many devices.

\begin{figure}[h]
	\centering
	\includegraphics[width = 0.55\linewidth]{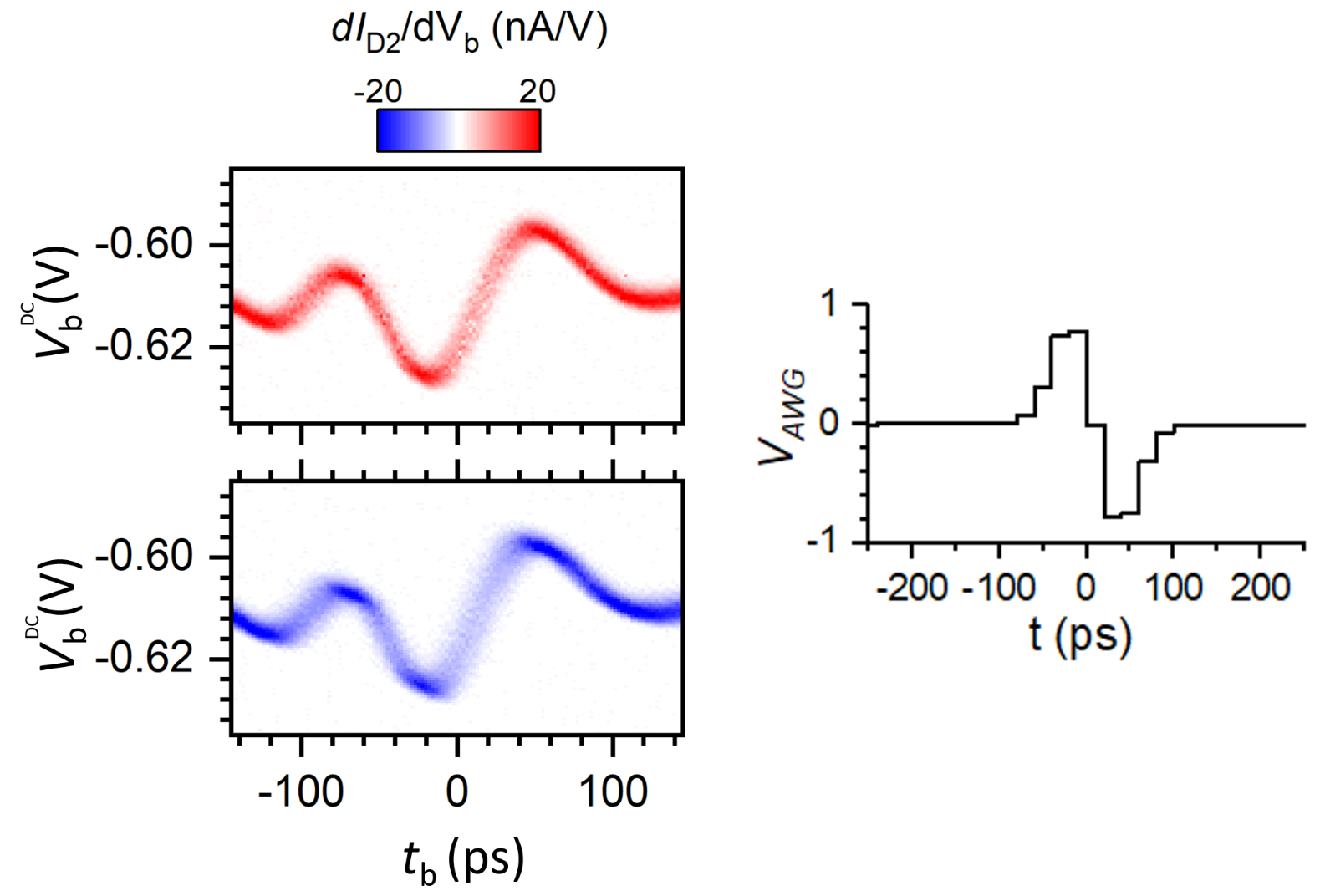}
	\caption{Maps of drain current derivative $dI_{\rm D2}^{\rm DC}/dV_{\rm b}$ versus $V_{\rm b}^{\rm DC}$ and $t_b$ for pump 1 and pump 2 at the point of synchronisation. The polarity of the derivative is related to which side of the barrier the electrons originate. The deviations from the programmed waveform (right hand side) are related to the finite bandwidth of the tramsmission line.
}
	\label{fig:time_matching}
\end{figure}

\subsection{Time matching}

Many factors govern the precise value of the phase delay required to achieve electron collision at the centre barrier. This includes RF cable lengths, ejection point within the pump waveform, travel time of the electron. As the interval of the pumping period where collision effects are seen is only $\simeq 5$~ps, an {\it in situ} synchronisation test is preferable to measuring these effects separately. To find the delay setting corresponding to $t_{12} = 0 $ a synchronisation pulse is applied to the centre barrier. This waveform is synchronised with both electron pumps but with an adjustable delay $t_b$ between the sync and pump signals. The DC threshold for blocking electrons is shifted by the instantaneous value any time-varying voltages at the moment of electron arrival.\cite{fletcher2013clock} These shifts enable synchronisation of the individual electron arrival times against the synchronisation signal, and hence against each other.\cite{Johnson2017} In practise, we adjust the pump-to-pump phase until the synchronisation pulses signatures are aligned along the sync-to-pump phase axis $t_b$ as shown in Fig.~\ref{fig:time_matching}. The ambipolar, short duration pulse (also shown in Fig.~\ref{fig:time_matching}) unambiguously identifies a particular phase range and avoids interference with electron loading processes that occur earlier in the cycle.

\begin{figure}[h]
	\centering
	\includegraphics[width = 0.6\linewidth]{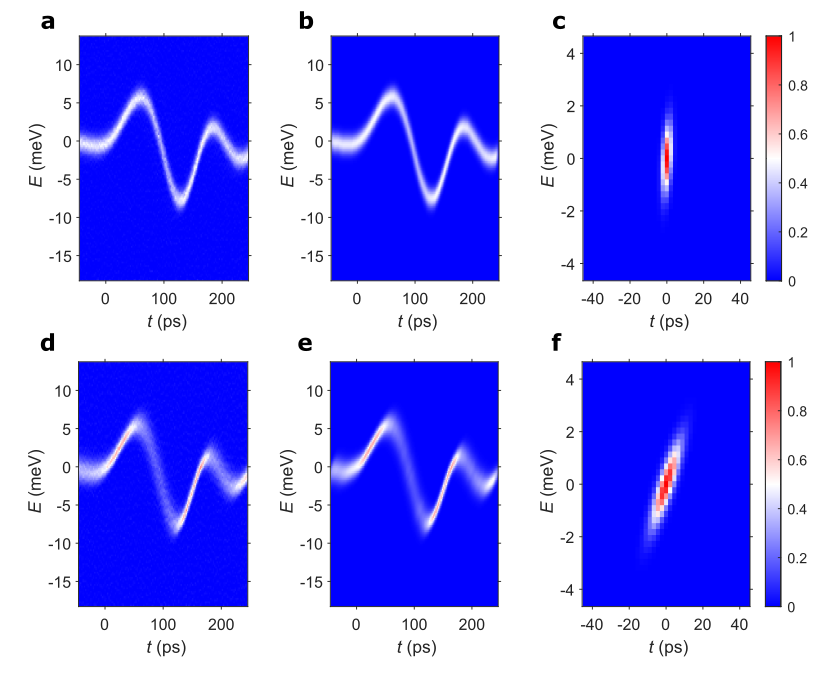}
	\caption{Estimating properties of input emission distributions. (a) Normalised synchronisation pulse as sampled with source S1 (as in Fig.\ref{fig:time_matching}) (b) Recreation of experimental data using bivariate model gaussian kernel shown in (c). (d-f) as above but for source S2.}
	\label{fig:wigner_distribution}
\end{figure}

\subsection{Estimation of emission energy-time functions}

Using modulated barriers in the electron path it is possible to perform measurements to reconstruct the electron energy-time distribution of these pumps.\cite{fletcher2019continuous} This reveals a variation in the emission energy and arrival time that are important for the electron collision. It has been shown that the emission distribution can be approximated by a bivariate gaussian distribution.\cite{fletcher2019continuous} This is characterised by an energy broadening $\sigma_E$, time broadening $\sigma_t$ and a coefficient $\rho$ which captures the energy-time correlation.\cite{Kashcheyevs2017} Here we approximately estimate these parameters for our device without a full tomography. This is based on data from the synchronisation pulse (see above) which is broadened by the finite emission energy-time distribution.\cite{fletcher2013clock} A parameterised model of this broadening is estimated using a convolution technique and values of $\sigma_t$, $\sigma_E$, $\rho$ can be found which reproduce the measured distribution. Figure \ref{fig:wigner_distribution}a-c show the sync data, synthesized distribution and estimated Wigner functions for source S1. Figure \ref{fig:wigner_distribution}d-f show the same for source S2. The parameters we extract are similar to those seen in other devices with the full tomography.\cite{fletcher2019continuous}

\begin{figure}[h]
	\centering
	\includegraphics[width = 0.55\linewidth]{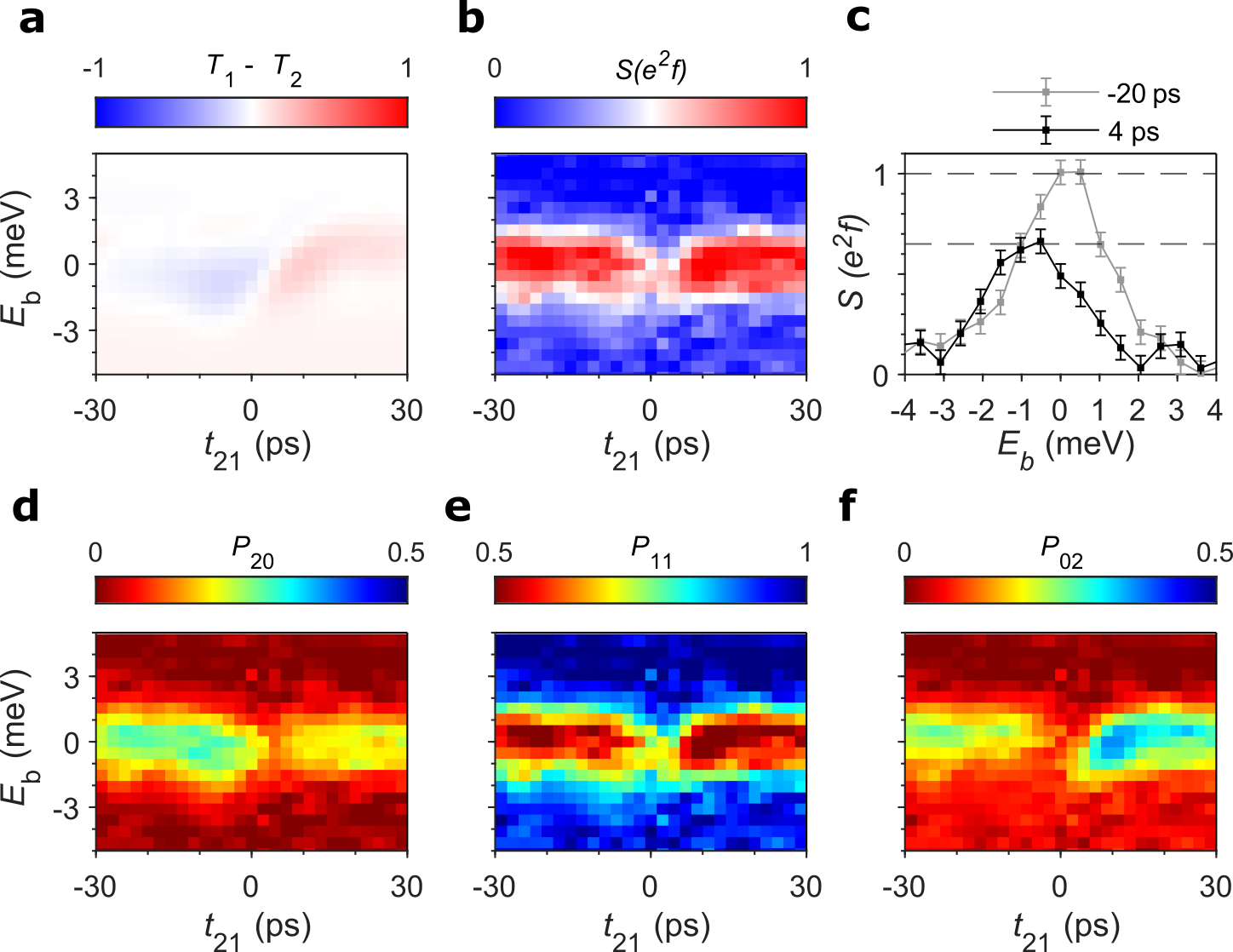}
	\caption{Noise and partioning data in additional sample: (a) Transmission imbalance $T_1 - T_2$ as a function of barrier height $E_b$ and time delay $t_{21}$. (b) Shot noise $S(E_b, t_{21})$. (c) Shot noise cuts far-from (grey) and close-to (black) synchronisation. (cd, (e) and (f) Partition probabilities  $P_{20}$, $P_{11}$ and $P_{02}$.}
	\label{fig:second_sample}
\end{figure}

\subsection{Factor affected visibility and reproducing partition statistics in a second sample}

A shift and suppression of the shot noise peak has been reproduced in three devices. Effects on the current transmission at the barrier was seen in these and several others where noise was not measured. Empirically we find that a requirement for visibility of the collision effect is that the pump emission time distribution is not too long ($\sigma_t << 10$~ps). Broader electron emission distributions show increasingly weaker collision effects. This was seen in both in devices with naturally broader emission distributions and also in the device reported in the main paper when emission was deliberately slowed.\cite{fletcher2019continuous} This dependence of the visibility upon the emission time is naturally captured in our model of ensemble averaging as timing fluctuations will naturally dilute the effects of the Coulomb interaction.

To test the reproducibility of more detailed aspects of the partioning effect we have we performed a run of comparable resolution on an additional sample nominally identical to the device shown in the paper. This was measured under equivalent conditions, i.e. $B = 10$~T, $T \simeq 0.1$~K, $f = 500$~MHz. Small additional technical differences include: use of a non-sinusoidal pump waveform to improve plateau flatness,\cite{giblin2012towards} use of a single noise amplifier to measuring the autocorrelation noise (with interleaved thermal background measurements subtracted), and small adjustments to account for cross-talk, as described below. The transmission imbalance $T_1 - T_2$ and partition noise $S$ for this device are shown in Fig.\ref{fig:second_sample}a,b. An imbalance of $T_1 - T_2$ is seen either side of $t_{21} \simeq 0$. This has the same sign as the data in the main paper i.e. transmission of early-arriving electron is favoured. A suppression and shift to lower $E_b$ of the noise peak is also seen, shown in Fig.\ref{fig:second_sample}c. When the partition statistics are computed, as in Fig~\ref{fig:second_sample}d-f, these display similar features to the data reported in the main text. This includes an enhancement $P_{11}|_{\rm min}>0.5$ and imbalances of $P_{02}$ and $P_{20}$.

\subsection{Exact energy matching and contributions to current imbalance}

Although the data in Fig.~3a are dominated by the `lobe' structure of current imbalance there are two small contributions to $T_1 - T_2$ that are not related to the coulomb interaction. Both are well understood. The first is related to the precise matching of the two input distributions. As described in the text, we are in the limit where the change in transmission of electrons at the barrier $T(E_{\rm b})$ is dominated by the size of the electron distribution, parameterised with $\sigma_E$. This means that, except for the case of precisely identical injection distributions (e.g. $\sigma_{E1} = \sigma_{E2}$), the transmission imbalance $T_1 - T_2$ plotted in Fig.~3a will show a contribution relating to the difference of the input distributions. This effect is included in the model via the parameters for the two Wigner functions which take the real, slightly mismatched, experimental values.

Additionally, we have observed that rotating the relative phase of the signal generators driving the two pumps can slightly `detune' the electron energies, shifting the value of $V_b$ at which each electron is blocked. These shifts have a periodic structure in $t_{21}$ on much longer time scales (200-1000~ps) than the small range of $t_{21}$ for the Coulomb interaction. The resulting shifts $E_{\rm CT} \lesssim \sigma_{En}$ over a $\simeq 60$~ps scan range window. One way to minimise this effect is to choose operating point where $dE_{\rm CT}/dt_{21} \rightarrow 0$. This technique was used to collect the data shown in Fig.3 in the main paper. Another technique is to characterise the shifts in the individually-measured values of $E_n(t_{21})$ and use small adjustments to the pump exit barrier height $\Delta V_{E}/V_{E} \leq 4 \times 10^{-3}$ to compensate for this at each value of $t_{21}$. This was used in the measurement in Fig.~\ref{fig:second_sample} above.

\section{Methods for computing the partition probabilities}
\label{sec:classical_tot}

We provide the method for calculating the partition probabilities discussed in the main text. The method is based on the trajectories of hot electrons ejected from the electron pump sources. Below, we discuss the validity and the details of this method.

\subsection{Parameter regime}
\label{sec:validity}

We summarize the experimental parameter regime and the validity of the method.
First, a characteristic energy scale of the barrier for wave packet collision is the energy window $\Delta_b$ of quantum scattering over which the barrier transmission probability changes from 0 to 1. In the parameter regime of $\Delta_b \ll \sigma_E$ (where $\sigma_E$ is the energy width of the electron wave packet), the most energy parts of the wave packets are either perfectly transmitted or perfectly reflected by the barrier, and these parts follow, and hence can be well described by, classical trajectories of the $\vec{E} \times \vec{B}$ drift along equipotential lines of the barrier as well as the Coulomb potential generated by the interaction between the packets injected to the barrier from the opposite sides;
in the regime, only the small energy portion of the packets in the window $\Delta_b$ undergoes quantum scattering by the barrier, and this portion is negligible~\cite{park2022arrival}. Second, recent tomographic measurements have shown \cite{fletcher2019continuous} that the purity of the wave packets generated by usual hot-electron pump sources is quite low. This indicates that the initial wave packets generated by the sources can be described by the ensemble of the initial conditions of classical particles having the electron charge.
Hence, the partition probabilities observed by our experiment can be analyzed by the ensemble average over the classical trajectories connecting from the sources to the detectors D1 or D2 depending on the initial conditions of the trajectories in the classical ensemble.

To estimate $\Delta_b$, we use our measurement data in another regime where  the electron energies $E_1$ and $E_2$ differ more than 4 meV. In this regime, the Coulomb interaction between the electrons is weaker than the regime considered in the main text, so the barrier energy scale $\Delta_b$ can be extracted from the measurement data. We estimate $\Delta_b \sim 0.1$ meV, by using the experimental values of
the window of the time delay $t_{21}$ (over which $I^\textrm{DC}_\textrm{D1}$ and $I^\textrm{DC}_\textrm{D2}$ change due to the synchronized arrival of single-electron packets) and the time uncertainties of the electron packets. The estimation details will be given in another paper.\cite{fletchermismatch}
And the energy width $\sigma_E$ of the two electrons is found as $\sigma_{E1} = 0.85$ meV and $\sigma_{E2} = 1.05$ meV (see Methods).
This implies that our experiment is in the regime of  $\Delta_b \ll \sigma_E$.

\subsection{Detailed theory}
\label{sec:classical}

We first explain the classical trajectories. The $\vec{E} \times \vec{B}$ drift motion along equipotential lines follows 
\begin{equation}
\dfrac{d \vec{r}_i}{dt} = \dfrac{1}{|B|^2} \vec{E}_i \times \vec{B} = \dfrac{1}{e |B|^2} \vec{\nabla}_i U_{\textrm{tot}} \times \vec{B},
\label{eq:velocity}
\end{equation}
where $\vec{r}_i(t) = (x_i(t), y_i(t))$ describes the trajectory of an electron generated by source $i=1,2$ that has its initial position $(x_{0,i}, y_{0,i})$ at time $t=0$. The electric field $\vec{E}_i = \vec{\nabla}_i U_{\textrm{tot}} /e$ is obtained from the potential $U_{\textrm{tot}}$,
and $- e$ is the electron charge. The potential $U_{\textrm{tot}}$ is decomposed into the electrostatic potential $U_{\textrm{2D}}(\vec{r}_i)$ of the barrier and the Coulomb repulsive interaction $U_{\textrm{ee}}(\vec{r}_1, \vec{r}_2)$ between electrons injected to the barrier from the opposite sides,
\begin{equation}
U_{\textrm{tot}}(\vec{r}_1,\vec{r}_2 ) = U_{\textrm{2D}}(\vec{r}_1) + U_{\textrm{2D}}(\vec{r}_2) + U_{ee}(\vec{r}_1, \vec{r}_2).
\end{equation} 
The barrier potential is modeled~\cite{fertig1987transmission} as 
the saddle constriction potential 
\begin{equation}
U_{\textrm{2D}}(x, y) = E_b - \frac{1}{2} m^* \omega_x^2 x^2 + \frac{1}{2} m^* \omega_y^2 y^2, 
\end{equation} 
where $m^*$ is the effective mass in GaAs, $E_b$ is the barrier height, and $\omega_x$ and $\omega_y$ are the frequencies describing the curvature of the barrier potential. The Coulomb interaction between two electrons at positions $\vec{r}_1$ and $\vec{r}_2$ is written as
\begin{equation}
U_{\textrm{ee}}(\vec{r}_1, \vec{r}_2) = \dfrac{e^2}{4\pi \epsilon_0 \epsilon_r} \left[ \dfrac{1}{|\vec{r}_1 - \vec{r}_2|} - \dfrac{1}{[ (\vec{r}_1 - \vec{r}_2)^2 + (2 d_{\textrm{2DEG}} )^2 ]^{1/2}} \right],
\end{equation}
where $\epsilon_0$ is the vacuum permittivity and $\epsilon_r = 12.9 $ is the relative permittivity of GaAs.
The second term of the above expression describes the effect of the screening~\cite{SkinnerPhysRevB.82.155111} by the top gate separated from the 2DEG by distance $d_{\textrm{2DEG}}$.

\begin{figure}[ht]
	\centering
	\includegraphics[width = 0.7\linewidth]{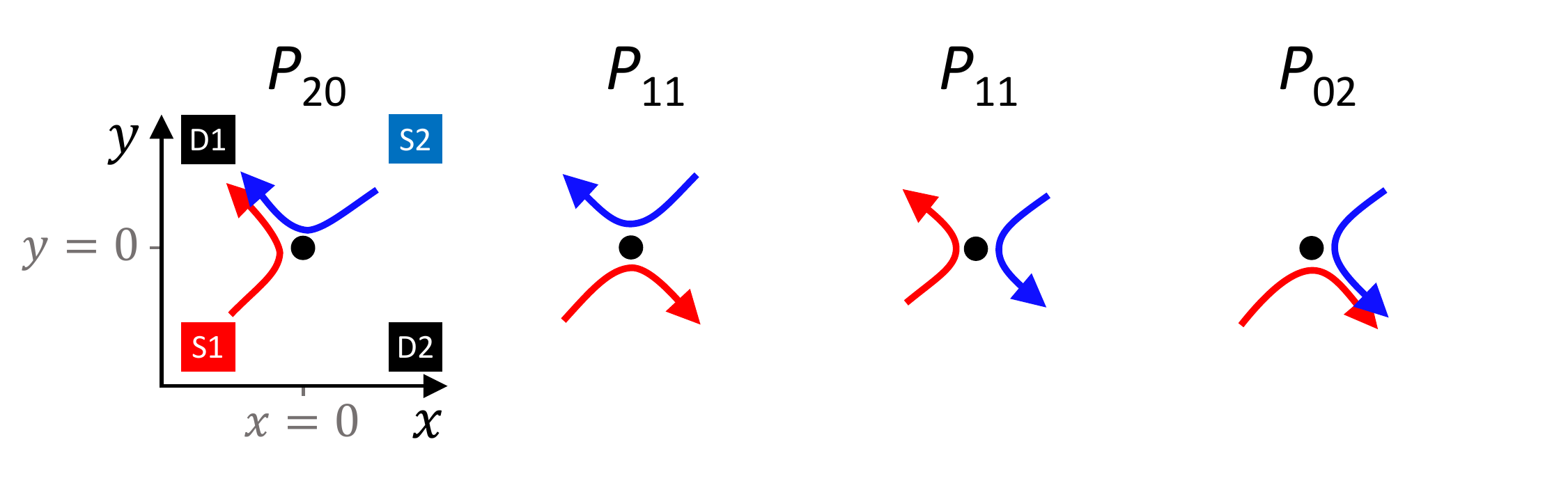}
	\caption{Trajectories of two electrons for the partition probabilities $P_{20}$, $P_{11}$, $P_{02}$.  The positions of detectors D1, D2 and sources S1, S2 are schematically shown in the $(x,y)$ coordinate. The $x$ and $y$ coordinates of source S1 are negative, while those of S2 are positive. The $x$ coordinate of detector D1 is negative, while that of D2 is positive. }
	\label{fig:partition_statistics}
\end{figure}

The ensemble of the initial positions $(x_{0,i}, y_{0,i})$ of the classical trajectories in the two-dimensional space is determined by the quasi-probability Wigner distribution $W_i (E_i, t_{a,i})$ obtained in our experiment (see Methods). Here the index $i=1,2$ stands for an electron generated by the source $i$ (see Fig.~\ref{fig:partition_statistics}). $E_i$ and $t_{a,i}$ are the energy of the electron and its arrival time at the position of detecting the Wigner distribution, respectively. We choose the initial positions such that, the initial distance between the two electrons is large enough to neglect their initial Coulomb interaction $U_{\rm{ee}}$. The initial positions $(x_{0,i}, y_{0,i})$ are located along equipotential lines of $E_i = U_{\textrm{2D}}(x_{0,i},y_{0,i})$. The relative location in the equipotential line is determined by the arrival time $t_{a,i}$; the electron having a larger arrival time is located farther away from the saddle point and arrives at the saddle point later time.
Note that the information about the initial momentums $(p_x, p_y)$ is not necessary, since the trajectories follow the $\vec{E} \times \vec{B}$ drift motion of the guiding center along equipotential lines in the strong perpendicular magnetic field.

We describe how to compute the partition probabilities $P_{n_\textrm{D1} n_\textrm{D2}}$, based on the classical trajectories $\big( x_i (t),y_i (t) \big)$ obtained by Eq.~\eqref{eq:velocity} with initial positions $(x_{0,i},y_{0,i})$. Here $(n_\textrm{D1}, n_\textrm{D2}) = (2,0)$, $(1,1)$, and $(0,2)$. The probabilities are obtained as
\begin{equation}
P_{20} = \int^{\infty}_{-\infty} d E_1  \int^{\infty}_{-\infty} d t_{a,1}  \int^{\infty}_{-\infty} d E_2  \int^{\infty}_{-\infty} d t_{a,2}  \, W_1(E_{1}, t_{a,1}) W_2(E_{2}, t_{a,2}) \lim_{t \to \infty}  \Theta( -x_{1}(t) ) \Theta( -x_{2}(t) ),
\label{eq:P20}
\end{equation}
\begin{equation}
P_{11} = \int^{\infty}_{-\infty} d E_1  \int^{\infty}_{-\infty} d t_{a,1}  \int^{\infty}_{-\infty} d E_2  \int^{\infty}_{-\infty} d t_{a,2}  \,  W_1(E_{1}, t_{a,1}) W_2(E_{2}, t_{a,2}) \lim_{t \to \infty} [ \Theta( x_{1}(t) ) \Theta( -x_{2}(t) ) + \Theta( -x_{1}(t) ) \Theta( x_{2}(t) ) ],
\label{eq:P11}
\end{equation}
\begin{equation}
P_{02}  = \int^{\infty}_{-\infty} d E_1  \int^{\infty}_{-\infty} d t_{a,1}  \int^{\infty}_{-\infty} d E_2  \int^{\infty}_{-\infty} d t_{a,2}  \, W_1(E_{1}, t_{a,1}) W_2(E_{2}, t_{a,2}) \lim_{t \to \infty}  \Theta( x_{1}(t) ) \Theta( x_{2}(t) ),
\label{eq:P02}
\end{equation} 
where $\Theta(x)$ is the Heaviside step function. In this calculation of the trajectories and the partition probabilities in Eqs.~\eqref{eq:P20}-\eqref{eq:P02}, the detector D1 is located at a position of a very large negative $x$, while the other detector D2 is at a position of a very large positive $x$ (see Fig.~\ref{fig:partition_statistics}).

\subsection{Connection with quantum theories}
\label{sec:quantum}

The classical calculations of Eqs.~\eqref{eq:P20}--\eqref{eq:P02} can be derived from the quantum mechanical calculations of the partition probabilities~\cite{park2022arrival}. We provide the key steps of the derivation.
First, the quantum mechanical Hamiltonian includes the single particle potential $U_\textrm{2D}$, the two-electron interaction $U_\textrm{ee}$, and the strong magnetic field. The Hamiltonian is decomposed into the part of the guiding center motion and the cyclotron motion~\cite{tsukada1976tail}. 
In the strong magnetic field, it is enough to follow the guiding center motion for computing the partition probabilities, since the radius of the cyclotron motion is sufficiently short.
Second, applying the Weyl transformation~\cite{polkovnikov2010phase} of the Hamiltonian part for the guiding center motion, we obtain a Hamiltonian function. The Hamilton equation of motion of this function is identical to the the classical evolution of the $\vec{E} \times \vec{B}$ drift in the strong magnetic field.
Here we use the fact that $U_{\textrm{tot}}$ is smooth in the length scale of the magnetic length $l_B = ( \hbar / e |B|)^{1/2}$, which is $\sim 8 \textrm{ nm}$ for $10 \textrm{ T}$.
Third, using the truncated Wigner approximation for the regime of $\Delta_b \ll \sigma_E$ and the Wigner-Weyl transformation of initial density matrices and observables~\cite{polkovnikov2010phase},  the partition probabilities are expressed in terms of the initial Wigner distribution and the classical trajectories.

We note that the results obtained from the above classical method in the regime of $\Delta_b \ll \sigma_E$ are also reproduced by another full quantum scattering theory~\cite{Sungguen2022} developed for a quantum wave packet of energy uncertainty $\sigma'_E \ll \Delta_b$. In this case, the initial classical ensemble of the energy uncertainty $\sigma_E$ is described by another classical ensemble $\mathcal{E}$ of quantum wave packets of energy uncertainty $\sigma'_E$, and the partition probabilities are obtained from the ensemble average of the quantum scattering probabilities over the classical ensemble $\mathcal{E}$.

\bibliography{Collision.bib} 

\bibliographystyle{apsrev}